\documentclass[]{article}
\usepackage{graphicx}

\title{Applications and a Three-dimensional Desktop Environment for an Immersive Virtual Reality System}

\date{}

\author{Akira Kageyama and Youhei Masada\\[1em]
Graduate School of System Informatics, \\
Kobe University, Kobe 657-8501, Japan}

\begin{document}
\maketitle

\begin{abstract}
We developed an application launcher called Multiverse
for scientific visualizations in a CAVE-type virtual reality (VR) system.
Multiverse can be regarded as a type of three-dimensional (3D) desktop environment.
In Multiverse,
a user in a CAVE room can
browse multiple visualization applications with 3D icons and
explore movies that float in the air.
Touching one of the movies causes ``teleportation'' into the application's VR space.
After analyzing the simulation data using the application,
the user can jump back into Multiverse's VR desktop environment in the CAVE.
\end{abstract}

\section{Introduction\label{sec:intro}}
CAVE is a room-sized virtual reality (VR) system, which was developed in the early 1990s
at the University of Illinois, Chicago~\cite{Cruz-neira1993}.
In a CAVE room, the viewer is surrounded by wall screens and a floor screen.
Stereo-images are projected onto the surfaces.
Tracking systems are used to capture
the viewer's head position and direction.
The wide viewing angle provided by the surrounding screens on the walls and floor
generates a high-quality immersive VR experience.
The viewer can interact with three-dimensional (3D) virtual objects
using a portable controller known as wand, in which the tracking system is installed.

CAVE systems have been used for scientific visualizations
from the first system~\cite{Cruz-neira1993}
until the latest generation (StarCAVE)~\cite{Defanti2009a}.
For example, visualization applications in CAVE systems have been
developed to analyze general computational fluid dynamics (CFD)~\cite{Jaswal1997},
turbulence simulations~\cite{Tufo1999},
CFD of molten iron~\cite{Fu2010},
CFD of wind turbines~\cite{Yan2011},
seismic simulation~\cite{Chopra2002},
meteorological simulation~\cite{Ziegeler2001},
biomedical fluid simulation~\cite{Forsberg2000},
magnetic resonance imaging~\cite{Zhang2001},
geomagnetic fields~\cite{Bidasaria2005},
archaeological studies~\cite{Acevedo2001}, and
geophysical surveys~\cite{Lin2011}.

%
\begin{figure}[h]
  \begin{minipage}{0.49\textwidth}
    \includegraphics[width=\textwidth]{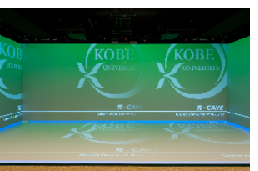}
    \caption{\label{fig:00}
      Overview of the $\pi$-CAVE system installed at Kobe University.
    }
  \end{minipage}
  \hspace{0.02\textwidth}
  \begin{minipage}{0.49\textwidth}
    \includegraphics[width=\textwidth]{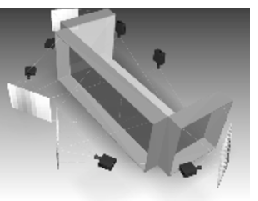}
    \caption{\label{fig:01}
      Projectors and mirrors used by the $\pi$-CAVE system.
    }
  \end{minipage}
\end{figure}
%

Recently, a new CAVE system was installed at the
Integrated Research Center (IRC) at Kobe University.
This CAVE system was named ``$\pi$-CAVE'' after the IRC's location on
Port Island (PI).
Fig.~\ref{fig:00} shows a front view of the $\pi$-CAVE while Fig.~\ref{fig:01} shows
the configuration of its projectors and mirrors.

The original CAVE system had a cubic geometry with a side length of 3 m.
A straightforward extension to enlarge the VR space of a CAVE is to
use a rectangular parallelepiped shape.
More sophisticated configurations have been proposed for advanced CAVE systems,
such as StarCAVE~\cite{Defanti2009a}, but
we used the rectangular parallelepiped approach for $\pi$-CAVE to
maximize the VR volume in the space allowed in the IRC building.
The side lengths of $\pi$-CAVE were 3 m $\times$ 3 m $\times$ 7.8 m.
As far as we know, this is the largest CAVE system in Japan.

We have developed several VR applications for the scientific
visualization of large-scale simulation data.
Of these,
Virtual LHD~\cite{Kageyama1998} was our first VR visualization application.
This application was developed for the CompleXcope CAVE system
installed at the National Institute for Fusion Science, Japan.
Currently, Virtual LHD is used to visualize the
magnetohydrodynamic (MHD) equilibrium state of a nuclear fusion experiment.
We also developed a general-purpose visualization application,
VFIVE~\cite{Kageyama2000,Ohno2007,Ohno2010,Ohno2010},
for 3D scalar/vector field data.
Recently, we added a new
visualization method to VFIVE at $\pi$-CAVE
for visualizing magnetic field lines frozen into a fluid~\cite{Murata2011a}.
The original VFIVE only accepted a structured grid data format as the input, but
an extension of VFIVE for unstructured grid data was
developed at Chuo University~\cite{Kashiyama}.
The development and its applications of VFIVE are summarized 
in our recent papers~\cite{Kageyama2013a,Kageyama2013b}.

In addition to improvements of VFIVE, we also developed the following
four types of novel CAVE visualization applications
for $\pi$-CAVE.
(1) IonJetEngine: for VR visualization of plasma particle in cell (PIC) simulations of an ion jet engine in space probes
(2) RetinaProtein: for molecular dynamics (MD) simulations of proteins
(3) SeismicWave: for the simulation of seismic wave propagation
 (4) CellDivision: to simulate three-dimensional
time sequence microscope images of mouse embryos.
All of these new CAVE visualization programs were
written using OpenGL and CAVElib.
We started developing these visualization applications when
the construction of $\pi$-CAVE was underway.

Several problems occur if multiple CAVE visualization applications are executed one after another, as follows.
First, the command has to be typed in to launch the first application
using the keyboard beside the CAVE room.
The user then enters the CAVE room wearing stereo glasses.
After analyzing the data from the first application in the CAVE,
the user leaves the CAVE room and takes off the glasses.
Next, the user types in the command to launch the second application
and enters the CAVE room wearing the stereo glasses.
These steps have to be repeated if there are many applications.
This inconvenience occurs because
the CAVE must be used for single tasks.

To resolve this inconvenience, we
developed an application launcher for CAVE.
This program, Multiverse, is a CAVE application written
in CAVElib and OpenGL.
Multiverse can control other VR applications.
These sub-applications are depicted in CAVE's VR space using 3D icons or panels.
If the user in the CAVE room touches one of the panels
using the wand,
they are ``teleported'' to the corresponding VR application.

In this paper, we report the hardware used by the $\pi$-CAVE system
in section~\ref{sec:picave}, and we describe the
design and implementation of Multiverse in section~\ref{sec:multiverse}.
The visualization applications loaded into Multiverse are
described in section~\ref{sec:applications}.

\section{$\pi$-CAVE system\label{sec:picave}}

\begin{figure}[h]
\begin{center}
 \includegraphics[width=0.5\textwidth]{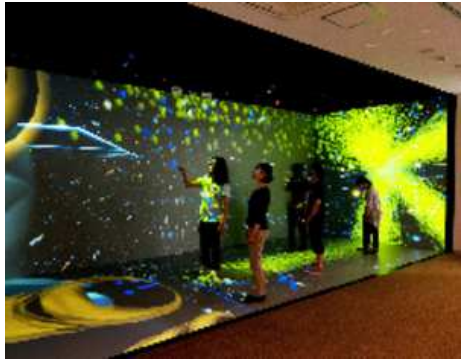}
    \caption{\label{fig:02}
     Alternative view of the $\pi$-CAVE system.
    }
\end{center}
\end{figure}

$\pi$-CAVE has a rectangular parallelepiped configuration with side
lengths of 3 m $\times$ 3 m $\times$ 7.8 m (Fig.~\ref{fig:02}).
The large width (7.8 m) is one of the characteristic features of the CAVE system.
The large volume of $\pi$-CAVE allows several people to stand on the floor at the same time, without any mutual occlusion of the screen views in the room.

%
%

\begin{figure}[h]
\begin{center}
 \includegraphics[width=0.5\textwidth]{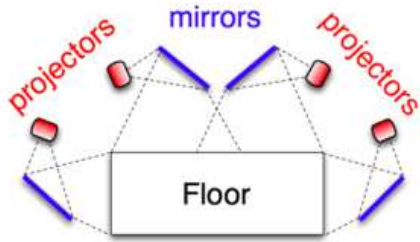}
    \caption{\label{fig:03}
     Projector settings for $\pi$-CAVE, viewed from above.
    }
\end{center}
\end{figure}

Like many other CAVE systems, $\pi$-CAVE has four screens: three wall screens (front, right, and left) and a floor screen.
Soft, semi-transparent screens are used on the walls.
The images are rear-projected onto these screens.
The floor is a hard screen where the stereo image is projected from the ceiling. Two projectors are used to generate the front wall image (Fig.~\ref{fig:03}). An edge blending technique is applied to the interface between the two images. Another pair of projectors is used for the floor screen. Each side wall screen (right and left) is projected onto using a projector.
In total, six projectors are used.

%
%

\begin{figure}[h]
\begin{center}
 \includegraphics[width=0.5\textwidth]{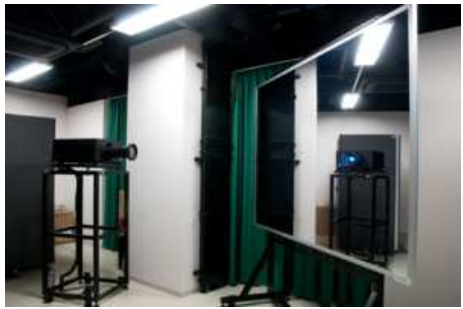}
    \caption{\label{fig:04}
     Pair of projectors and mirrors.
    }
\end{center}
\end{figure}

The resolution of the projector (Christie WU12K-M) shown in Fig.~\ref{fig:04}
with the counterpart mirror,
is 1920 $\times$ 1200 pixels. The brightness is 10,500 lumens.
An optical motion tracking system (Vicon) is used for head and wand tracking.
Ten cameras with 640 $\times$ 480 resolution are installed on top of the wall screens.
A commonly used API (Trackd) is used for the interface to CAVElib.

%
%

Two computer systems are used for computations and for rendering $\pi$-CAVE.
One is a Linux PC (HP Z800) with 192 GB of shared memory.
Three sets of GPUs (NVIDIA QuadroPLEX) are used for real-time
stereoscopic image generation by the six projectors.
The other computer system is a Windows PC cluster system.

We used OpenGL for the graphics API and CAVElib for the VR API.
We are also aiming to use VR Juggler~\cite{Bierbaum2001} for the VR API.
Some of our first trials using VR Juggler can be found in our report~\cite{Meno2012}.

\section{Multiverse\label{sec:multiverse}}

\begin{figure}[h]
\begin{center}
 \includegraphics[width=0.5\textwidth]{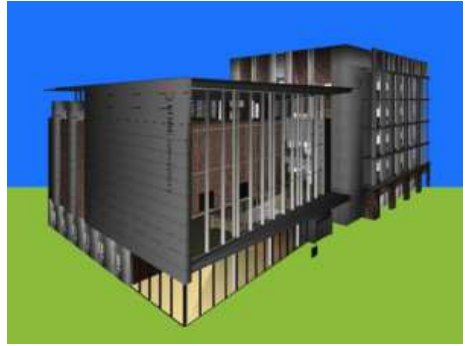}
    \caption{\label{fig:05}
     Three-dimensional CAD data for the IRC building loaded in Multiverse.
    }
\end{center}
\end{figure}

\begin{figure}[h]
\begin{center}
 \includegraphics[width=0.6\textwidth]{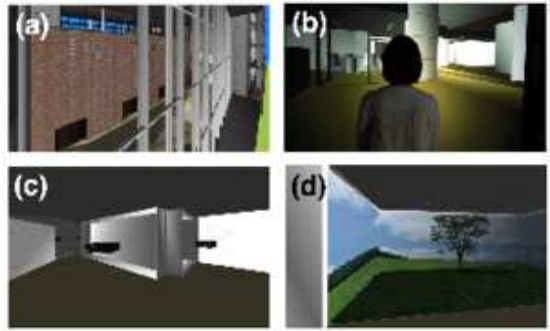}
    \caption{\label{fig:06}
     A snapshot sequence of Multiverse in the World mode.
     (a) The viewer is entering the virtual IRC building.
     (b) The viewer walks into the building.
     (c) Rear of the (virtual) $\pi$-CAVE.
     The CAD models of the projectors behind the CAVE screens can be seen.
     (d) Virtual $\pi$-CAVE in the real $\pi$-CAVE.
    }
\end{center}
\end{figure}

We developed an applications launcher, Multiverse, for the $\pi$-CAVE system.
At the start of this Multiverse environment,
the viewer in the $\pi$-CAVE
stands in the virtual building in IRC where $\pi$-CAVE is installed.
The 3D CAD model data of the IRC building (Fig.~\ref{fig:05}) is loaded into
Multiverse and rendered in 3D in real time.
This is the Multiverse's start-up environment known as \textit{World}.
In the \textit{World} mode of Multiverse, the viewer can
walk through the building.
Fig. ~\ref{fig:06}(a) shows a snapshot where the user is approaching
the IRC building.
In Fig.~\ref{fig:06}(b), the viewer is (literally) walking into the (virtual) IRC building.
Some fine structures
of the building, including the virtual $\pi$-CAVE
is shown in Fig.~\ref{fig:06}(c) and~(d), are also loaded from CAD data files.

%
%

%
%

In Multiverse, there are two methods
of showing the application list loaded in Multiverse.
The first is to use ``ribbons'' that connect the wand and application icons.
In the ``ribbons'' mode,
the user in the \textit{World}
finds one or more curves or wires that start from the wand tip.
Each wire is a type of guide that leads the user to a Gate.

\begin{figure}[h]
\begin{center}
 \includegraphics[width=0.5\textwidth]{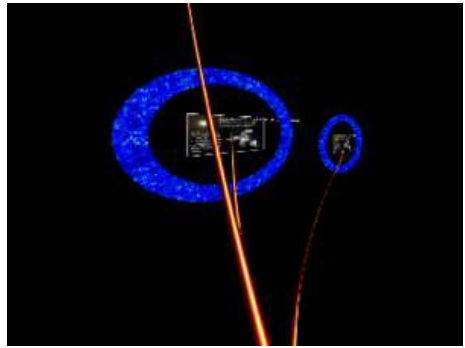}
    \caption{\label{fig:06b}
     Gates to the individual visualization applications floating in the \textit{World}.
    }
\end{center}
\end{figure}

A Gate is an entrance to the VR world of the corresponding application.
If multiple visualization applications are loaded into Multiverse,
this automatically generates the corresponding number of Gates.
All of these are connected to the user (or the wand) via guide wires (Fig.~\ref{fig:06b}).
If the user walks or ``flies'' into a place in front of a Gate,
they will find an exploratory movie near the Gate
(see the rectangular panel in the center of the blue, torus-shaped Gate in Fig.~\ref{fig:06b}).
This explains the type of application that will be executed when the user selects the Gate.
To select the application, the user (literally) walks through the Gate when the
corresponding VR application program loads and the user feels as
if they have been ``teleported'' to the visualization space.
Each VR world is known as a \textit{Universe} in Multiverse.

%
%

\begin{figure}[h]
\begin{center}
 \includegraphics[width=0.5\textwidth]{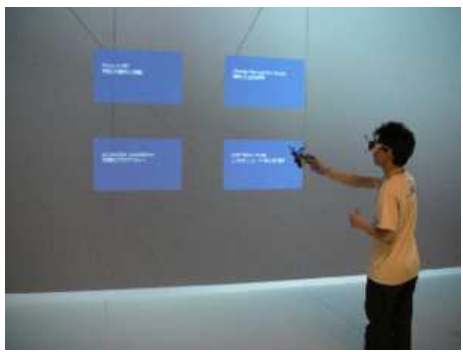}
    \caption{\label{fig:07}
     Virtual touch screens in the \textit{World} mode.
     Each panel represents a VR visualization program or \textit{Universe}.
     A \textit{Universe} is executed when the user touches the panel using the wand.
    }
\end{center}
\end{figure}

Another method of showing the application lists loaded in the Multiverse is to
use a virtual elevator.
When the user enters the elevator in the (virtual) IRC building,
they are automatically taken upward by the elevator into the sky above the IRC building.
The spatial scale of the view changes rapidly from the building, to the
city, country, and finally the globe.
The user finds that they are ``floating'' in space surrounded by stars.
Several panels then appear in front of the viewer.
Each panel represents a visualization application (Fig.~\ref{fig:07}).

%
%
When the user touches one of the panels,
the corresponding VR application is launched
and the user is ``teleported'' to the selected visualization \textit{Universe}.

In short, Multiverse is composed of the \textit{World} and several \textit{Universes}.
\textit{World} is a type of 3D desktop environment
and a \textit{Universe} is a visualization application loaded onto Multiverse.

In the program code, each \textit{Universe} is simply a standard CAVE application with a unified interface to the Multiverse class.
A \textit{Universe} is an instance of a class that is derived from a virtual class known as Vacuum. Vacuum represents an empty space, which only has an interface to the Multiverse class through the member functions
\verb+initialize()+, \verb+draw()+, \verb+update_per_frame()+, and \verb+compute()+.
These function names convey their roles to readers who are familiar with CAVElib programming.

\section{Applications\label{sec:applications}}

In this section,
we describe five applications, or \textit{Universes}, which we
developed as the first applications for the Multiverse environment.

\subsection{Universe::GeomagField}
We converted VFIVE, which is described in section~\ref{sec:intro}, into a class of \textit{Universe}.
VFIVE is a general-purpose visualization tool,
so we can visualize any vector/scalar field provided that the data are legitimate for VFIVE's input data format in the Multiverse framework.

%
\begin{figure}[h]
\begin{center}
 \includegraphics[width=0.5\textwidth]{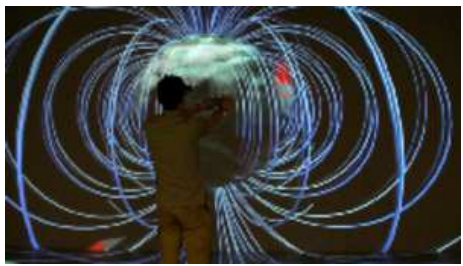}
    \caption{\label{fig:08}
     GeomagField, a \textit{Universe} in Multiverse.
     This VR application was used to visualize an MHD simulation of
     geomagnetic field generation in Earth's liquid core.
     }
\end{center}
\end{figure}
%

Fig.~\ref{fig:08} shows a snapshot of an example of a \textit{Universe} based on VFIVE, known as GeomagField. The input data used by GeomagField was a geodynamo simulation performed by one of the authors and his colleagues~\cite{Kageyama2008,Miyagoshi2010,Miyagoshi2011}.
The purpose of this simulation was to understand the mechanism that generates the Earth's magnetic field (or geomagnetic field).

%
\begin{figure}[h]
\begin{center}
 \includegraphics[width=0.5\textwidth]{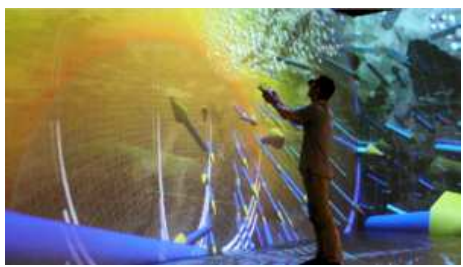}
    \caption{\label{fig:09}
      Another snapshot of GeomagField.
      The scalar field distribution of the temperature is visualized by volume rendering.
      The convection flow velocity is visualized using
      3D arrow glyphs while tracer particles are under the spotlight.
     }
\end{center}
\end{figure}
%

Fig.~\ref{fig:09} shows another snapshot of GeomagField in which
two VFIVE visualization methods were applied.
The temperature distribution was visualized by volume rendering (colored in orange to yellow).
The 3D arrow glyphs show the flow velocity vectors around the wand position.
The arrows followed the motion when the viewer moved the wand, which
changed the directions and lengths (vector amplitudes) in real time.
The white balls are tracer particles that also visualized the flow velocity.
These balls were highlighted in a spotlight or cone-shaped region, the apex of which was the wand.
This visualization method is known as Snowflakes in VFIVE.
The viewer can change the focus of the flow visualization by
changing the direction of the spotlight via wand direction movements.

\subsection{Universe::IonJetEngine}

\begin{figure}[h]
\begin{center}
 \includegraphics[width=0.5\textwidth]{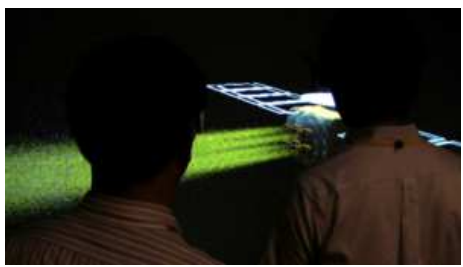}
    \caption{\label{fig:10}
      A snapshot of the Universe known as IonJetEngine.
      Plasma jets from a space probe were visualized by moving ions (yellow) and
      electrons (blue) as particles. The simulation data were provided by Prof.~Usui.
     }
\end{center}
\end{figure}

The second example of a \textit{Universe} is known as IonJetEngine
and a snapshot is shown in Fig.~\ref{fig:10}.

%
%

This \textit{Universe} visualized a plasma PIC simulation of the ion jet engine of a space probe. The positions of the particles (ions and electrons) were represented by balls (yellow for ions and blue for electrons).
The velocity distribution of the jet was visualized as the set of the individual motions of the particles.
A 3D model of the virtual space probe from which the plasma jet beams were ejected is also shown in Fig.~\ref{fig:10}.

\subsection{Universe::RetinaProtein}

\begin{figure}[h]
\begin{center}
 \includegraphics[width=0.5\textwidth]{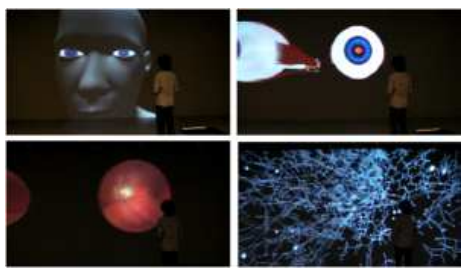}
    \caption{\label{fig:11}
      Snapshots of the RetinaProtein Universe.
      The molecular structure of rhodopsin was visualized in the human retina.
      The MD simulation data were provided by
      Prof.~Ten-no of Kobe University and his colleagues.
     }
\end{center}
\end{figure}

Fig.~\ref{fig:11} shows a \textit{Universe} known as RetinaProtein, which was a molecular dynamics simulation of rhodopsin~\cite{Akinaga2011},
a protein in the human retina. At the start of this \textit{Universe}, the viewer observed a 3D model of a human (see the top panel of Fig.~\ref{fig:11}). As the viewer approached the model's face, the fine structures of the eyes became visible until MD simulation visualization appeared.

%
%

\subsection{Universe::SeismicWave}

\begin{figure}[bh]
\begin{center}
 \includegraphics[width=0.5\textwidth]{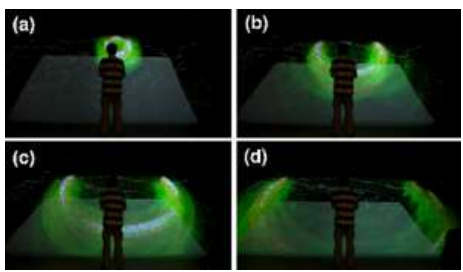}
    \caption{\label{fig:12}
      Time sequence of snapshots of the SeismicWave \textit{Universe}.
     }
\end{center}
\end{figure}

In this \textit{Universe}, a simulation of seismic wave propagation~\cite{Furumura2003}was visualized, which was performed by Prof.~Furumura of the University of Tokyo by animated volume rendering (see Fig.~\ref{fig:12}). In this \textit{Universe}, we implemented rapid volume rendering based on the 3D texture mapping technique in CAVEs.
The full details of this implementation will be reported elsewhere.

%
%
	
\subsection{Universe::CellDivision}
\begin{figure}[h]
\begin{center}
 \includegraphics[width=0.5\textwidth]{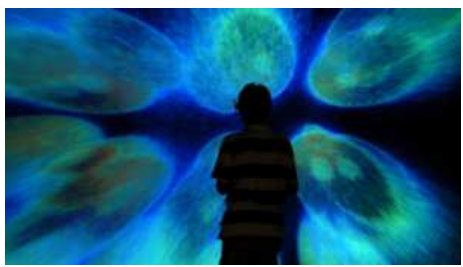}
    \caption{\label{fig:13}
      A snapshot of CellDivision.
      Animated volume rendering of live cell images of a mouse embryo.
      The data were provided by Dr.~K.~Yamagata, Osaka University.
     }
\end{center}
\end{figure}

The final \textit{Universe} described here is CellDivision and a snapshot is shown in Fig.~\ref{fig:13}.
The target data used for this visualization were not simulation data. Instead, they were microscope images of live mouse embryos.
The data were provided by Dr.~Yamagata of Osaka University.
The time sequence of microscope images was visualized as an animated volume rendering using the same tool used for SeismicWave in the previous subsection.

%
%

\section{Summary\label{sec:summary}}

In many CAVE systems, VR applications are executed
as single tasks.
Thus, the user has to type in each command one after another outside the CAVE room.
To convert a CAVE into a more convenient tool for scientific visualization,
we developed an application launcher known as Multiverse.
Multiverse comprises a \textit{World} and \textit{Universes}.
\textit{World}, which correspond to the desktop of a PC operating system,
where the user can select visualization applications by touching icons floating in the \textit{World}. Using the virtual touch screen interface, the specified application program is launched and the user is ``teleported'' to another VR space containing the corresponding visualization application, which is known as a \textit{Universe}.
We developed five \textit{Universes}, which can be launched from
the Multiverse environment.
Multiverse was designed as a general application framework, so
it can read and control other \textit{Universes}.
A user can jump back to a \textit{World} and switch
to another \textit{Universe} at any time from any \textit{Universe}.


During the implementation of Multiverse, we developed several new fundamental tools and methods for the CAVE environment, such as a fast speed volume renderer, a 3D model (CAD) data loader/renderer, and a 2D movie file loader/renderer. Details of these fundamental tools and methods will be reported elsewhere.

\section*{Acknowledgements}

We thank the undergraduate students at our laboratory at Kobe University (Toshiaki Morimoto, Yasuhiro Nishida, Yuta Ohno, Tomoki Yamada, and Mana Yuki) for contributing to the development of Multiverse. The plasma particle simulation data were provided by Prof.~H. Usui, Dr.~Y.~Miyake,
and Mr.~A. Hashimoto (Kobe University). The MD simulation data were provided by Prof.~S.~Ten-no and Dr.~Y. Akinaga.
The simulation data for seismic wave propagation were provided by Prof.~T.~Furumura (University of Tokyo). The microscope images were provided by Dr.~K.~Yamagata (Osaka University).

This work was supported by JSPS KAKENHI Grant Numbers 23340128 and 30590608,
and also by the Takahashi Industrial and Economic Research Foundation.

%
%
%


\providecommand{\newblock}{}

\end{document}